\documentclass[12pt,preprint]{aastex}

\slugcomment{Not to appear in Nonlearned J., 45.}

\shorttitle{A 4--6~GHz Spectral Scan of TMC-1}
\shortauthors{Kalenskii et al.}

\begin{document}


\title{A 4--6~GHz Spectral Scan and 8--10~GHz Observations
of the Dark Cloud TMC-1}


\author{S. V. Kalenskii}
\affil{Astro Space Center, Profsoyuznaya 84/32 117997 Moscow, Russia}
\email{kalensky@asc.rssi.ru}

\author{V. I. Slysh}
\affil{Arecibo Observatory, Arecibo, PR, USA}
\affil{Astro Space Center, Profsoyuznaya 84/32 117997 Moscow, Russia}
\email{vslysh@asc.rssi.ru}

\author{P. F. Goldsmith}
\affil{Department of Astronomy, Cornell University, and National Astronomy
and Ionosphere Center, Ithaca, NY 14853, USA}
\email{goldsmit@astrosun.astro.cornell.edu}

\and

\author{L. E. B. Johansson}
\affil{Onsala Space Observatory. SE-439 92 Onsala, Sweden}
\email{leb@oso.chalmers.se}

\begin{abstract}
The results of the lowest frequency spectral survey carried out toward
a molecular cloud and sensitive observations at selected frequencies
are presented. The entire Arecibo C-band (4--6~GHz) was observed towards
the cyanopolyyne peak of TMC-1 with an rms sensitivity of about
17--18 mK (about 2--2.5~mJy). In addition, a number of selected frequency ranges
within the C-band and X-band (8--10~GHz) were observed with longer
integration times and rms sensitivities  7--8 mK ($\approx$ 2 mJy)
or higher.  In the spectral scan itself, already--known
H$_2$CO and HC$_5$N lines were detected. However, in more sensitive
observations at selected frequencies, lines of C$_2$S, C$_3$S, C$_4$H, C$_4$H$_2$, HC$_3$N
and its $^{13}$C substituted isotopic species, HC$_5$N, HC$_7$N, and HC$_9$N
were found, about half of them detected for the first time.
The rotational temperatures of the detected molecules fall in the range
4--9 K. Cyanopolyyne column densities vary from $5.6\times 10^{13}$~cm$^{-2}$ for
HC$_5$N to $2.7\times 10^{12}$~cm$^{-2}$ for HC$_9$N.
Our results show that for molecular observations at low frequencies (4--10 GHz) to be useful
for studying dark clouds, the sensitivity must be of the order
of 5--10 mK or better. To date, observations at around 10~GHz have been more productive
than those at lower frequencies.

\end{abstract}




\section{Introduction}

Most known interstellar molecules have been detected in the millimeter
wavelength range. This is clearly
related to the fact that, in general, the simplest and hence the lightest
molecules are the most abundant cosmic molecular species.
Their strongest rotational lines at the temperatures characteristic of
the molecular interstellar medium arise at mm and submm wavelengths.
Heavier molecules, often with large permanent dipole moments, have detectable
rotational lines at lower frequencies in the microwave range. 
Transitions between low energy levels are more favorable in terms of line
intensities, especially in regions of
low or modest density and low temperature.
The latter include, in particular, the inner parts of dark clouds, which are among
the coldest regions of the interstellar medium, because they have no internal 
energy sources and are shielded
from external sources of radiation by their own gas and dust.

Another reason why observations at low frequencies are needed is the fact
that the line intensities of molecules having fractional abundance below approximately
$10^{-11}$ relative to H$_2$ fall below the confusion level in the millimeter 
wavelength range.  In such cases, 
sensitive observations at low frequencies, which are less ``contaminated''
by the emission of simple molecules, may help.
In addition, hyperfine splitting is often negligible at high frequencies, but
detectable at low frequencies (e.g. for species HC$_5$N, HC$_7$N, etc.).
Hyperfine splitting  
may be used to determine line opacities and thus to obtain more accurate 
molecular column densities and abundances.

Spectral scans, covering wide ranges of frequencies, are powerful tools
for molecular searches. Typically many lines of the same molecule fall
within the observed spectral window, making it possible to more reliably
identify observed spectral features. A number of spectral scans
of molecular clouds have so far been carried out in atmospheric windows
between 17~GHz and 700~GHz.  A selection of the frequency ranges 
covered and references are 
17.6--22 GHz \citep{bel93}; 
72--144 GHz  \citep{cum86,joh84,tur91};
330--360 GHz \citep{jew89}; 
455--507 GHz, \citep{whi03};
607--725 GHz, \citep{sch01}; 
780--900 GHz, \citep{com02}.
Recently, \cite{kai04} carried out a spectral scan of the dark cloud TMC-1
in the frequency range between 8.8 and 50 GHz. However, no spectral scan
has been performed at frequencies below 8.8~GHz.
Therefore we undertook a spectral survey of 
TMC-1 at 4--6 GHz. This is the lowest frequency spectral survey
carried out toward a molecular cloud\footnote{A similar spectral scan
of IRC+10216 has been performed using the Arecibo telescope
by Araya et al. (2003).}.

TMC-1 has a shape of a ridge elongated in the northwest-southeast direction.
The structure and chemical composition of TMC-1 have been studied by
\cite{hir92}, \cite{pra97}, \cite{tur00}, and \cite{dic01}.
The temperature of this object is about 10~K,
which is typical for dark clouds; the H$_2$ density estimates towards the
``cyanopolyyne peak'' (see below) vary between $4\times 10^3$~cm$^{-3}$
\citep{tur00} and $2\times 10^4$~cm$^{-3}$ \citep{pra97}.  The study by Peng et al.~(1998)
revealed 45 clumps which are grouped into 3 cylindrical features
oriented along the ridge. The LSR velocities of these features are approximately
5.7, 5.9, and 6.1~km~s$^{-1}$. TMC-1 has proven to be an excellent object 
for the study of chemistry in dark clouds under quiescent conditions.
It shows a carbon--rich chemistry with chemical gradients across the densest
part of the ridge extending over 0.2 pc x 0.6 pc.
The most prominent chemical feature of
TMC-1 is the high abundance of various carbon--chain molecules, such as 
cyanopolyynes HC$_{2n+1}$N and radicals C$_n$H.
The peak position of the line intensities of cyanopolyynes
(the cyanopolyyne peak) was found to be located $\approx 7'$ southeast from
another prominent position in TMC-1, the peak of the ammonia line emission.
The heaviest known interstellar
molecule, HC$_{11}$N, as well as a number of other heavy molecules
have been found exactly towards the cyanopolyyne peak, making it an especially
interesting object for observations at low frequencies.

\section{Observations}

The observations were performed with the 305-m radio telescope
of the Arecibo Observatory. The entire Arecibo C-band (4--6~GHz) was
observed towards the cyanopolyyne peak position
($\alpha_{1950}=04^h38^m38^s$, $\delta_{1950}=25^\circ35'45''$).
The C-band observations were performed in total power
mode which is effective as a result of the excellent system stability, and
which maximizes the signal to noise ratio.  
Two senses of linear polarization were independently observed and the two
outputs averaged together. The four correlator boards of
the spectrometer were offset in frequency by 5.25~MHz, and each was set
to a bandwidth of 6.25~MHz, thus yielding an instantaneous continuous
frequency coverage of 21~MHz with a frequency resolution of 3~kHz
(0.183 km~s$^{-1}$ at 5~GHz). Each frequency setup was observed for
5 minutes. The entire C-band was observed 3 times, with one observation
shifted by 3~MHz, yielding an rms sensitivity of about 17--18 mK
(about 2--2.3~mJy).

In addition, a number of selected frequency ranges
within the C-band and X-band (8--10~GHz) were observed with longer
integration times and rms sensitivities 7--8 mK 
or higher. The X-band observations were performed in two senses
of circular polarization. The beam size was about $0.'9 \times 1.'0$ in 
C-band and about $0.'5 \times 0.'5$ in X-band.
The gain was about 8 K/Jy in C-band and about 4~K/Jy in X-band.
The frequency ranges were centered at the frequencies of different
molecular transitions taken from the JPL catalog of line positions
and intensities (http://spec.jpl.nasa.gov/ftp/pub/catalog/catform.html)
and the Cologne database for molecular spectroscopy
(http://www.ph1.uni-koeln.de/vorhersagen/),
as well as from the catalog of known molecular lines by F.~Lovas
(http://physics.nist.gov/cgi-bin/micro/table5/start.pl).

Each scan of the total power observations consisted of 50 6--second
ON SOURCE observations followed by two 10--second calibration scans.
The first of these calibration scans was taken with a
noise diode switched on, and the second with the noise diode off.
The data were calibrated by dividing the ON SOURCE spectra
by the calibration spectra; the latter were calculated as the difference
of the two calibration signal scans. To optimize the accuracy of
the calibration while minimizing the effect of noise in the
calibration scans, the calibration scans were approximated by
polynomials of order 20 to 29. Polynomoials of low order (up to 3) were
applied to remove residual baselines from the calibrated spectra.  
Both the calibration and subsequent data reduction were performed 
with the CLASS software package.

Lines with resolved hyperfine structure (i.e., when the hyperfine
components are presented as different entries in Table~\ref{gauss})
were fitted using the standard CLASS method HFS. This method enables us
to derive the optical depth of the strongest component from
the relative intensities of the observed components
assuming that the excitation temperatures, LSR velocities, and linewidths
of individual components are the same.

\section{Results and analysis}

\begin{figure}
\plotone{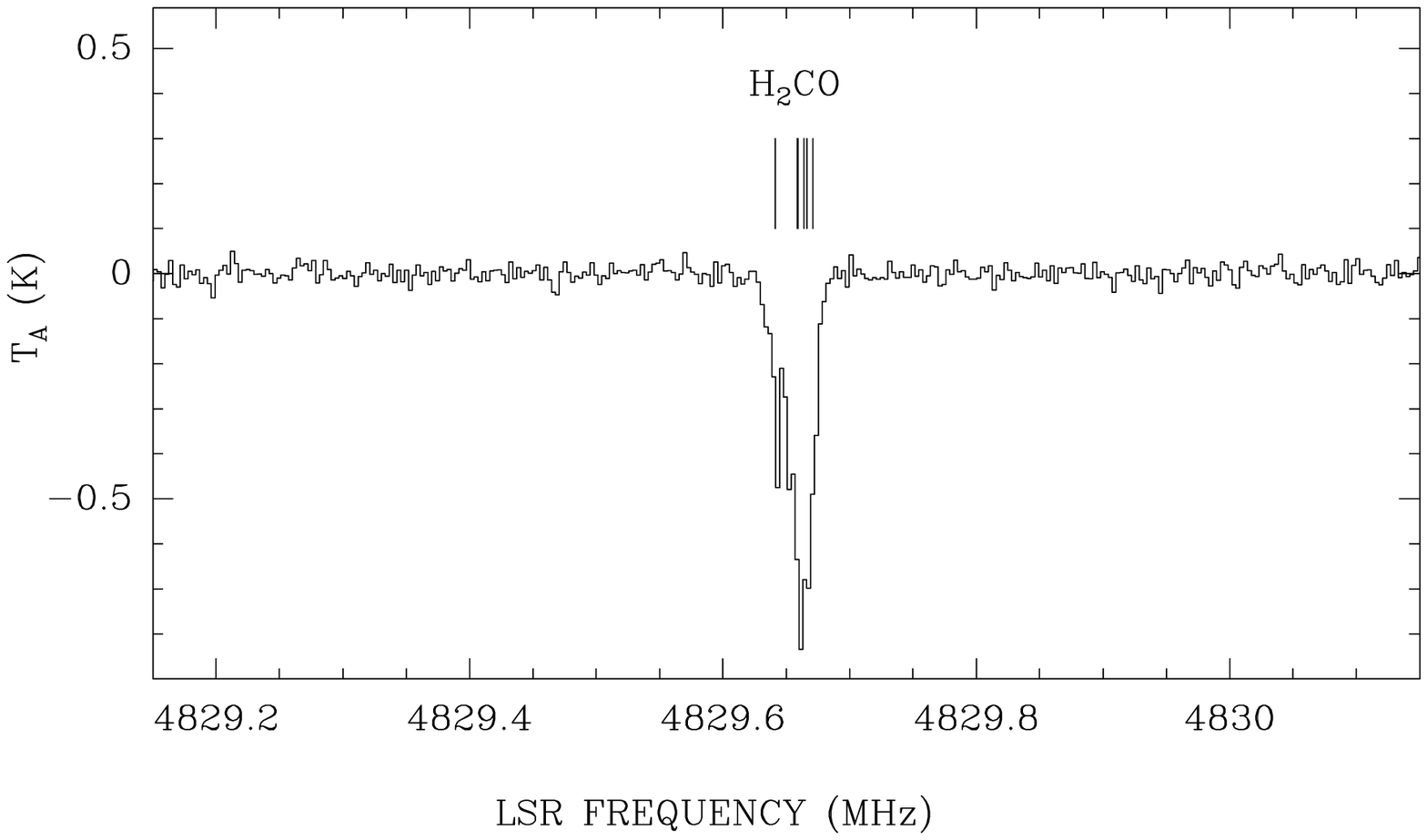}
\caption{A 1~MHz portion of the C-band spectrum. The horizontal axis is
the rest frequency in MHz and the vertical axis
is the antenna temperature in Kelvins. The strong absorption line
at the center is the $1_{10}-1_{11}$ transition of H$_2$CO. The positions
of the hyperfine components are shown by vertical bars.
\label{scan}}
\end{figure}

TMC-1 has a very narrow lines of less than 0.4~km~s$^{-1}$ width. In order
to achieve high velocity resolution of 0.18~km~s$^{-1}$ with the available
number of spectrometer channels the entire band of 4--6~GHz was divided
into 381 spectra. A portion of one of the C--band spectrum is shown in
Fig.~\ref{scan}\footnote{All spectra are available at
http://tanatos.asc.rssi.ru/~kalensky/scan.html}. As a result of
the spectral scan itself, we detected only lines of H$_2$CO and HC$_5$N. 
However, in more sensitive observations at selected
frequencies in C-- and X--bands we detected lines of C$_2$S, C$_3$S, C$_4$H,
C$_4$H$_2$, HC$_3$N and its $^{13}$C substituted isotopic species, HC$_5$N,
HC$_7$N, and HC$_9$N. The $J=5-4$ line of HC$_7$N and all lines of HC$_9$N,
C$_4$H$_2$, CCS, and C$_3$S are new, i.e., they were not included
in the catalog of known interstellar lines by F.~Lovas at the time of
writing this paper. The list of the detected lines with their fitted Gaussian
parameters is presented in Table~\ref{gauss}.

Most of our results were analysed with rotational diagrams. To construct
rotational diagrams, we derived a rotational level
population from the total intensity of relevant hyperfine components.

For optically thin emission, the integrated line intensity, $W$ (K cm s$^{-1})$,
is related to the upper level column density $N_u$ by the well-known relation:

\begin{equation}
\frac{N_u}{g_u} = \frac{3kW}{8\pi^3 \nu_0 S \mu^2}\;
\frac{J(\nu ,T_{ex})}{J(\nu ,T_{ex}) - J(\nu ,T_{bg})}~,
\end{equation}
where $S$ is the line strength, $\mu$ is the permanent electric dipole moment,
and $J(\nu,T) = h\nu/k \cdot [e^{(h\nu/kT)}-1]^{-1}$.  At low frequencies,
where $h\nu \ll kT$, $J(T,\nu)$ is approximately equal to $T$. 
$T_{ex}$ and $T_{bg}$ are the excitation temperature and the background brightness 
temperature, respectively.  If $J(\nu,T_{ex}) \gg J(\nu,T_{bg})$, the term
$[J(\nu ,T_{ex})]/[J(\nu ,T_{ex}) - J(\nu ,T_{bg})]$ approaches unity. 
In this case, assuming LTE and applying Boltzmann's equation, one can
obtain from Eq. (1):

\begin{equation}
\ln\frac{3kW}{8\pi^3 \nu_0 S \mu^2} = \ln\frac{N}{Q_{rot}} -
\frac{E_u}{kT_{rot}}~,
\end{equation}
where $E_u$ is the energy of the upper level of the transition, 
$Q_{rot}$ is the partition function, and $T_{rot}$ is the excitation temperature
characterizing the rotational transitions of interest.
Using Eq. (2), one can obtain the total molecular column density $N$ and rotational
temperature $T_{rot}$ from observations of several lines of the same
molecule. However, in the case of HC$_5$N and some other molecules
the rotational temperature is fairly low; hence, the assumption
$J(T_{ex},\nu) \gg J(T_{bg},\nu)$ is not valid, and the
term $[J(\nu ,T_{ex})]/[J(\nu ,T_{ex}) - J(\nu ,T_{bg})]$
must be taken into account. Therefore we derived $N$ and $T_{rot}$ applying
the following iterative approach:
\begin{enumerate}
\item
Determination of ``initial guesses'' of $T_{rot}$ and $N$ from Eq. (2)
\item
Calculation of $[J(\nu ,T_{rot})]/[J(\nu ,T_{rot}) - J(\nu ,T_{bg})]$
from the estimate of $T_{rot}$
\item
Determination of the next estimate of $T_{rot}$ and $N$ using
the calculated value of \\$[J(\nu ,T_{rot})]/[J(\nu ,T_{rot}) - J(\nu ,T_{bg})]$ 
\end{enumerate}

Steps 2 and 3 were repeated until convergence was achieved.    
The calculated rotational temperatures and column densities are given in
Table~\ref{trot}

For C$_4$H and C$_4$H$_2$, no rotational diagrams were constructed.
To estimate column densities of these species, we assumed that the energy
level populations are thermalized at a temperature taken from the literature and
presented in Table~\ref{trot}. First, the optical depths of the observed
lines were determined from the relation
$\tau = T_R/(J(T_{ex},\nu)- J(T_{bg},\nu))$, where $T_R$ is the main beam
brightness temperature. Then, using the prescription from the secton
``INTENSITY UNITS AND CONVENTIONS'' of the JPL catalog documentation
\break (http://spec.jpl.nasa.gov/ftp/pub/catalog/doc/catdoc.pdf)
absorption coefficients per molecule per km~s$^{-1}$, $\sigma_{ba}$, were found.
Column densities $N$ were obtained using the relation $\tau = \sigma_{ba}\;N$.
The same technique was applied to obtain the upper limits to the column density 
presented in Table~\ref{neg}.

\subsection {HC$_3$N, H$^{13}$CCCN, HC$^{13}$CCN, and HCC$^{13}$CN.}

\begin{figure}
\plotone{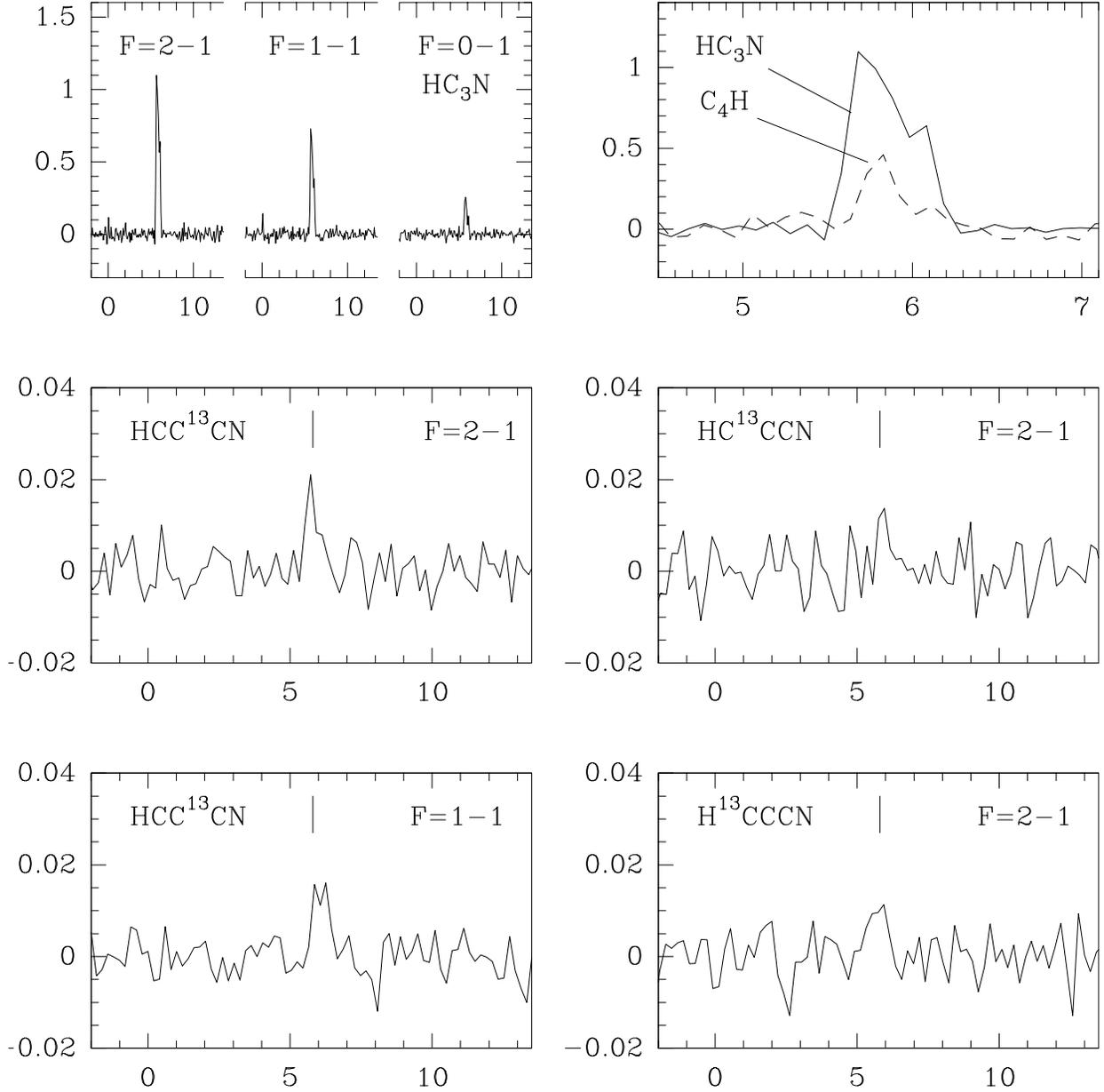}
\caption{Spectra of the $J=1-0$ transitions of HC$_3$N  
and $^{13}$C substituted isotopic species.
The horizontal axis is the LSR velocity in km~s$^{-1}$ and the vertical
axis is the antenna temperature in Kelvins. The C$_4$H line at 9497.616 MHz
is indicated by the dashed line in the plot at the upper right, together
with the $F=2-1$ line of HC$_3$N (solid line).
\label{hc3n}}
\end{figure}

The spectra of the $J = 1-0$ HC$_3$N emission are shown in Fig~\ref{hc3n}.
One sees a complex line profile, already found in a number of
other molecular lines, and explained in terms of the existence 
of small-scale structure in TMC-1 \citep{dic01, pen98}. Following
\cite{pen98}, who distinguished three main kinematic components
towards TMC-1, we approximated the hyperfine--split line by a sum of
three kinematic components with their intensities, LSR velocities,
and linewidths free. 
The LSR velocities of the components proved to be 5.69, 5.85, and
6.06~km~s$^{-1}$, which are approximately coincident with the velocity
components determined by Peng et al. and by Dickens et al.
A very limited number of offset measurements were
performed which showed that (1) the peak of the 5.69~km~s$^{-1}$ emission
is located approximately 40 arcsec south; (2) 
the peak of the 5.85~km~s$^{-1}$ emission
is located approximately 30 arcsec southeast;
and (3) the peak of the 6.06~km~s$^{-1}$ emission
is located approximately 40 arcsec north of the nominal position.
However, both the extent and the sampling of our ``map'' are insufficient
to determine the spatial distribution of the HC$_3$N emission.

The optical depth of the main hyperfine component is fairly small
(about 0.2 for the 5.69 and 5.85~km~s$^{-1}$ components); however, 
it is in agreement with optical depths about unity for the higher frequency
transitions reported by, e.g.,~\cite{sor86}. 
Therefore, simple methods of analysis,
such as rotation diagrams, are not adequate (the formal use of
the basic rotation diagram method yields a rotational temperature of
2 K), and a statistical equilibrium (SE) analysis which allows for
effects of finite optical depth should be used
\citep[see e.g.][]{gold99}.

In addition to the main species, 
we detected the $J=1-0$ HCC$^{13}$CN emission and marginally detected 
two other $^{13}$C substituted isotopomers, HC$^{13}$CCN
and H$^{13}$CCCN. The weakness of the lines of the $^{13}$C--subsituted species
did not allow us to measure reliably the isotopic
abundance ratios, but the results seem to be in a qualitative agreement with
the previous results of \cite{tak98}, obtained from the observations
of the $J=2-1$, $4-3$, and $5-4$ lines at higher frequencies, i.e.,
the HCC$^{13}$CN emission is the strongest and the HC$^{13}$CCN and
H$^{13}$CCCN emission is weaker by a factor of 1.3. 
The ratio of the HC$_3$N/HCC$^{13}$CN line intensities, $54\pm 15$ is
also in agreement with the ratio of HC$_3$N/HCC$^{13}$CN abundances of
55 found by Takano et al., indicating a small optical depth for the lines
observed using Arecibo.

\subsection {HC$_5$N.} 

\begin{figure}
\plotone{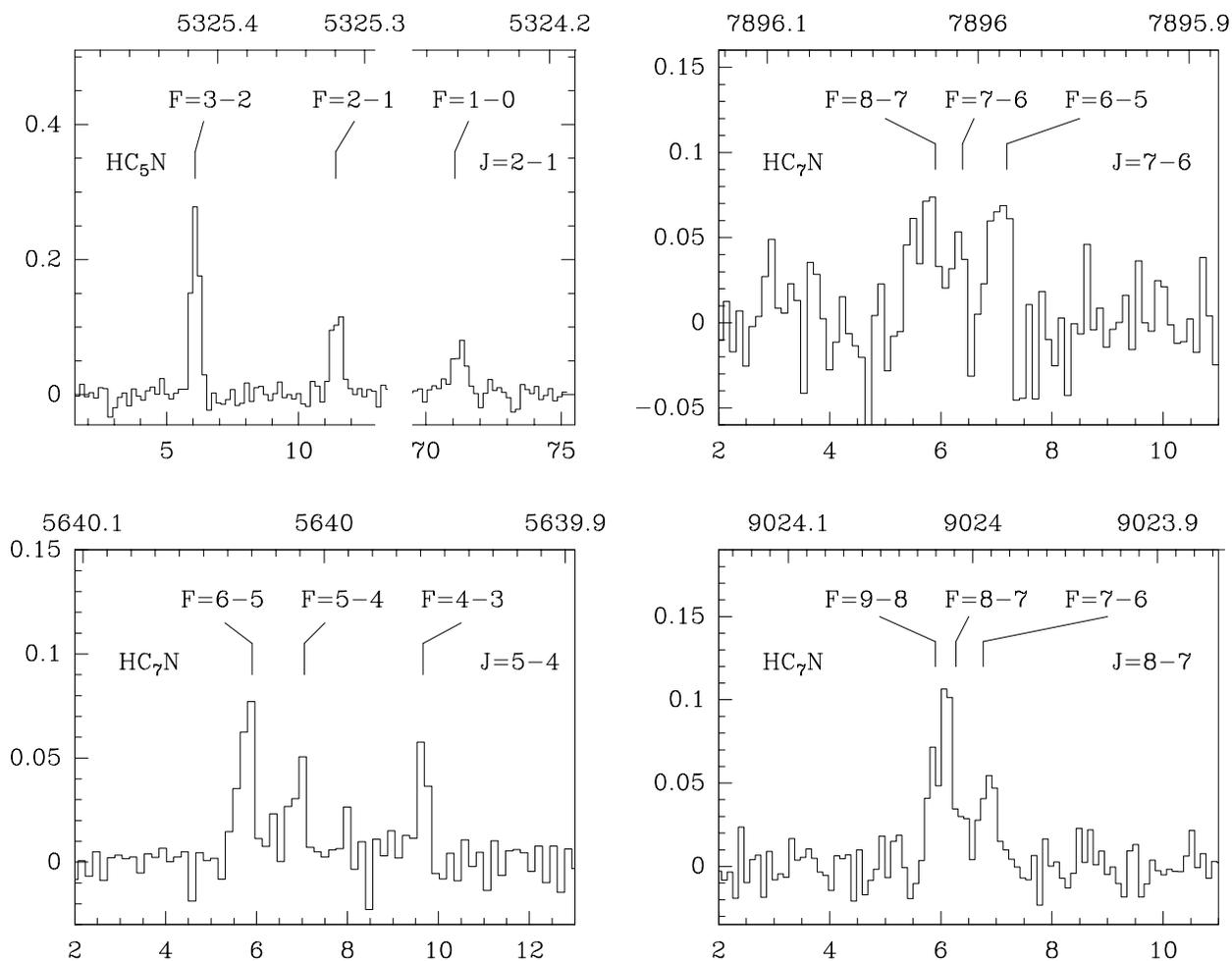}
\caption{Spectra of cyanopolyynes HC$_5$N and HC$_7$N. The rotational
transition is indicated in the upper right corner of each spectrum.
The vertical lines indicate the positions of the hyperfine components. 
The upper horizontal axis plots the rest frequency in MHz and the lower
horizontal axis is the LSR velocity of the strongest hyperfine
component in km~s$^{-1}$; the vertical axis is the same as in Fig.~2.
\label{cyanop}}
\end{figure}

We detected $J = 2-1$ emission of HC$_5$N (Fig.~\ref{cyanop}).
The ratio between the intensities of different hyperfine components
corresponds to LTE with small optical depth.
Combining our line intensity with the intensities of different lines of HC$_5$N
published by \cite{sne81}, in a rotational diagram (Fig.~\ref{rot}) we
derive a HC$_5$N rotational temperature
of 4.3~K, in agreement with e.g., \cite{bel98}, and a column density 
of $5.6\times 10^{13}$ cm$^{-2}$.

\subsection{HC$_7$N.}

The $J = 5-4$, $7-6$, and $8-7$ lines were observed and spectra are shown
in Fig. 3.  The rotation
diagram (Fig.~\ref{rot}) was constructed assuming HC$_7$N source size of
$6'\times 1.'3$~\citep{bel98}, and yields a rotational 
temperature of 7.7~K and a HC$_7$N column density of
$1.2\times 10^{13}$~cm$^{-2}$.

\subsection {HC$_9$N.}

\begin{figure}
\vspace{-170mm}
\plotone{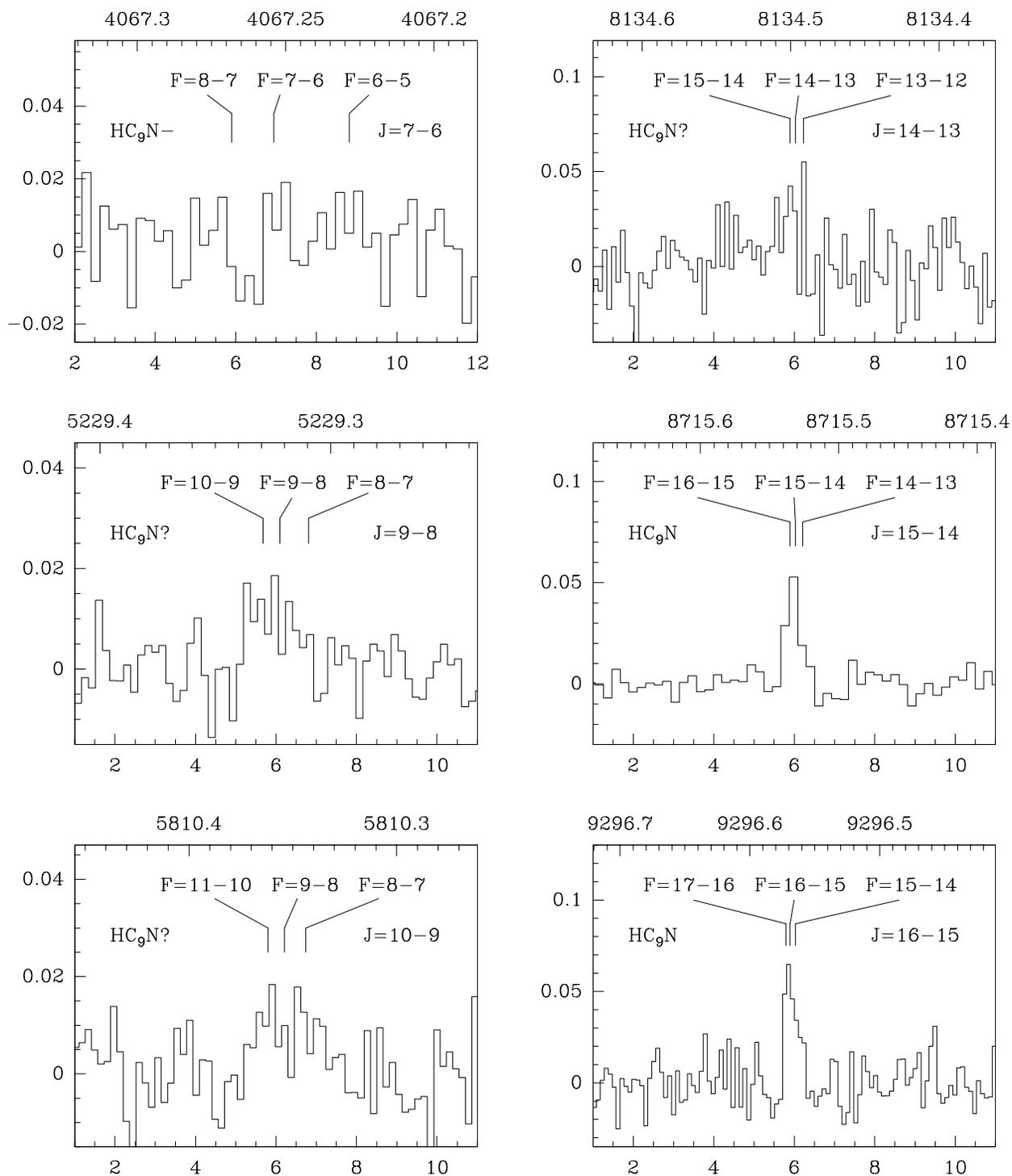}
\caption{Various HC$_9$N lines. The rotational
transition is indicated in the upper right corner of each spectrum.
The vertical lines indicate the positions of the hyperfine components.
A minus after the species name means that the line was not detected.
A question mark after the species name denotes a marginal detection.
The axes are the same as in Fig.~3.
\label{hc9n}}
\end{figure}

We detected the $J=15-14$ and $16-15$ transitions of HC$_9$N, 
and marginally detected the $J = 9-8$, $10-9$, and $14-13$ lines.
The $J=7-6$ line was not found because of the weakness of
the rotational transition and its hyperfine splitting (Fig.~\ref{hc9n}).
Rotational diagrams were constructed assuming a source size equal to
$6'\times 1.'3$ (Fig.~\ref{rot}, upper right) and $100''\times 55''$, suggested by
\cite{bel98} (Fig.~\ref{rot}, lower right). One can see that 
the best fit to our data is obtained for the former source size.
Thus, we derived a rotational temperature equal to 8.4~K and a HC$_9$N
column density equal to $2.7\times 10^{12}$~cm$^{-2}$.

Our results on cyanopolyyne rotational temperatures are in agreement
with the previous results by \cite{bel98}, who found that
the rotational temperatures of the cyanopolyynes increase with the number
of heavy atoms over the range HC$_5$N--HC$_9$N and explained this effect
in terms of less efficient radiative decay in the longer cyanopolyyne chains.

\subsection{Other detected molecules} 
\begin{figure}
\plotone{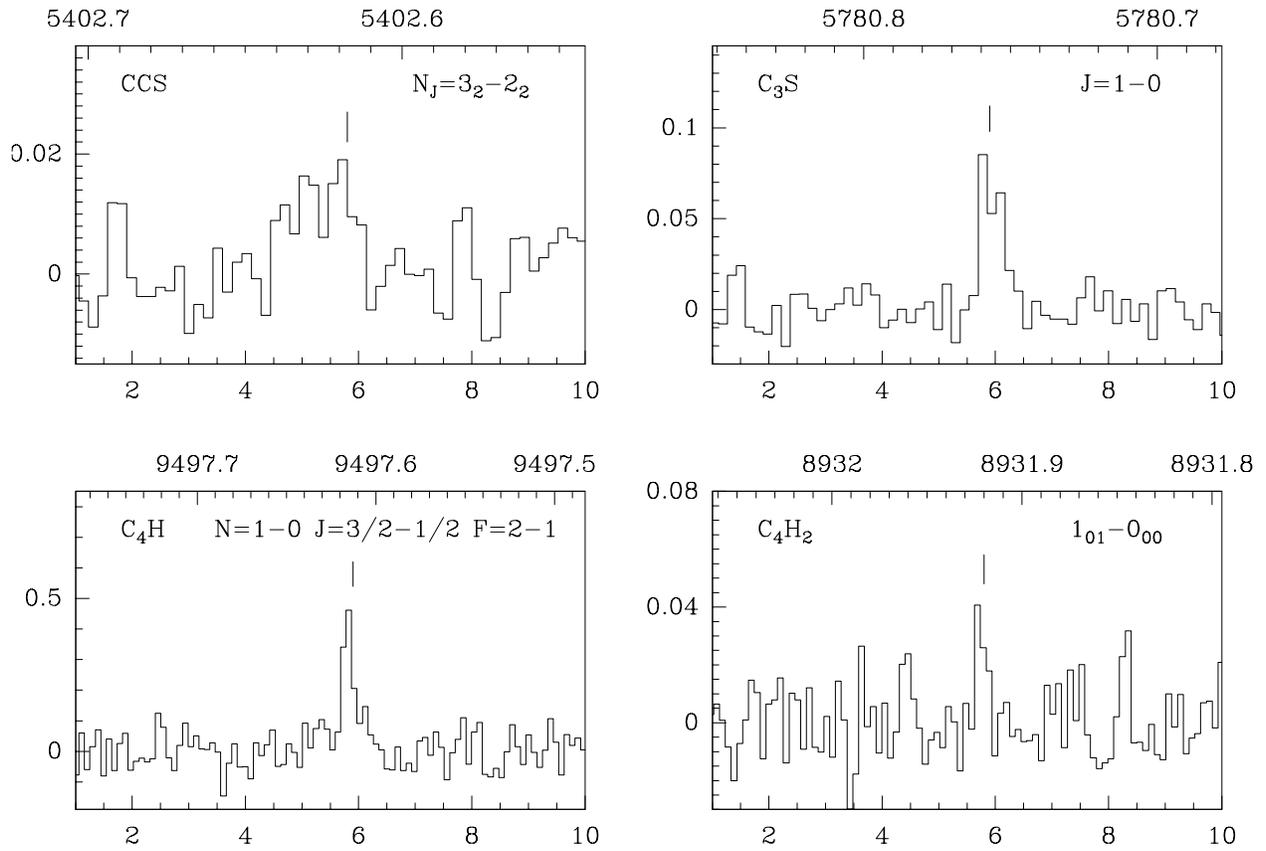}
\caption{Spectra of CCS, C$_3$S, C$_4$H, and C$_4$H$_2$ transitions.
The upper horizontal axis plots the rest frequency in MHz and the lower
horizontal axis is the LSR velocity in km~s$^{-1}$; the vertical axis is the same
as in Fig.~2.
\label{other}}
\end{figure}

\begin{figure}
\plotone{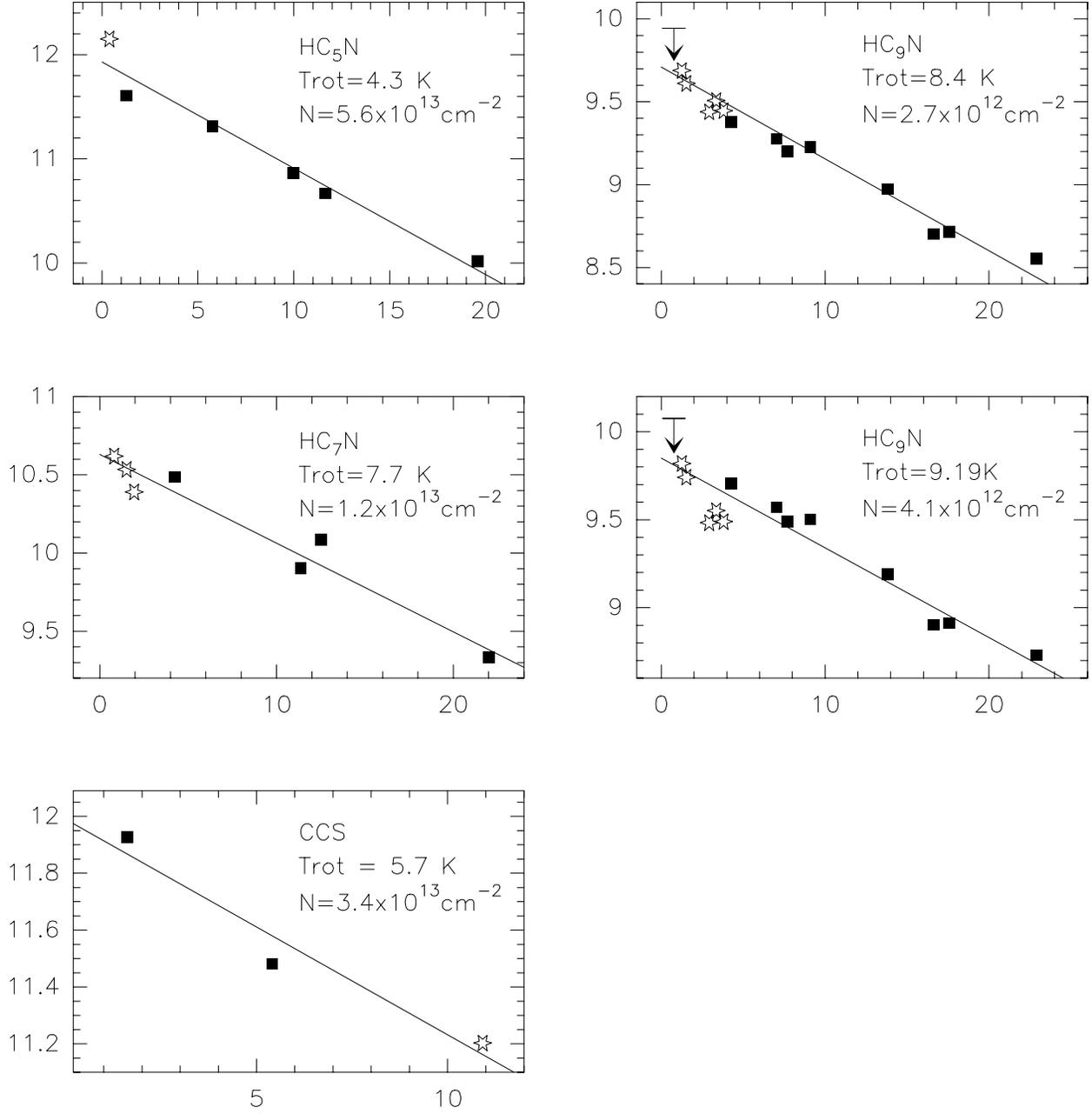}
\caption{Rotation diagrams. The lines detected at Arecibo are denoted
by stars, while filled squares represent the lines taken from the literature.
Two different diagrams for HC$_9$N were constructed assuming
different source sizes: $6'\times 1.'3$ (upper), and $1.7'\times 0.'9$
(lower). 
For each diagram, the horizontal axis is $E_u/k$ and the vertical axis is
$\log (3kW)/(8\pi^3 \nu_0 S \mu^2)$ (see section 3).
\label{rot}}
\end{figure}

An unexpected result is the
weakness of the tentatively detected $N_J=3_2-2_2$ line of CCS at
5402.6~MHz relative to the 1--0 line of C$_3$S at 5780.8~MHz
(Fig.~\ref{other}). The CCS abundance is higher by an order of
magnitude \citep{fue90}, and usually CCS lines are much stronger
than C$_3$S lines at approximately the same frequencies (e.g.,
\cite{dic01}). The CCS line is intrinsically weaker than the C$_3$S line
($S\cdot \mu^2 = 13.76$ for the C$_3$S line and only 6.71 for the CCS line).
The relevant rotational diagram (Fig.~\ref{rot}) yields
a rotational temperature of 5.7~K and a CCS column density of
$3.4\times 10^{13}$~cm$^{-2}$, which is in good agreement with
that derived by means of a LVG analysis by \cite{fue90}. 
We derived the C$_3$S column density from LVG calculations, assuming
$T_{kin}=10$~K and n$_{H_2}=3\times 10^4$~cm$^{-3}$ \citep{pra97}.
Our value, $3\times 10^{12}$~cm$^{-2}$, is in good agreement with
that derived by Fuente et al. Thus, the weakness of the $3_2-2_2$
line of CCS is a consequence of its intrinsic weakness, the high
location of its energy levels ($E/k$ equals 10.84~K for the $N_J=3_2$ CCS
level), and the fairly low rotational temperature of CCS.

We detected a strong $N=1-0$, $J=3/2-1/2$, $F=2-1$ transition of C$_4$H  
(Fig.~\ref{hc3n}), already observed in TMC-1 by Bell et al.~(1982, 1983).
The Arecibo line is narrow as a result of the weakness of
the 6.06~km~s$^{-1}$ component towards the observed
position. Since this component is fairly strong in the $1-0$ line of HC$_3$N,
our data suggest that there is a difference between the small scale (about $30''$)
distributions of HC$_3$N and C$_4$H. Note that the same  C$_4$H line,
observed with a much larger $3'.4$ beam (see Fig.~2 in \cite{bel82}),
is broader, having a width of about 0.4~km~$s^{-1}$, which is typical of TMC--1. 

We believe that the detection of the $1_{01}-0_{00}$ para-C$_4$H$_2$
line is real, since the C$_4$H$_2$ column density (derived from
the line intensity under the assumptions that the C$_4$H$_2$
rotational temperature is 4.2~K and the ortho-to-para ratio is
4.2 \citep{kaw91}) is equal to $4.6\times 10^{12}$~cm$^{-2}$, 
which is comparable to, but somewhat smaller than, the value of 
$7.5\times 10^{12}$~cm$^{-2}$ derived by Kawaguchi et al.

The $1_{10}-1_{11}$ H$_2$CO line, detected using Arecibo (Fig.~\ref{scan};
Table~\ref{gauss}) is rather similar but not identical to that observed
previously by e.g., \cite{hen81} with a $2'.6$ beam. The
relative intensities of hyperfine components yielded the optical
depth of the main component to be 1.6 and the line excitation
temperature to be 1.2~K.
No $1_{10}-1_{11}$ H$_2^{13}$CO and H$_2$C$^{18}$O lines were detected.
We estimate the lower limits to the ratios of the $1_{10}$ level populations
for H$_2$CO/H$_2^{13}$CO and H$_2$CO/H$_2$C$^{18}$O assuming
equal excitation temperatures for the $1_{10}-1_{11}$
transition of all isotopic species. 
The limits are 24 and 76, which are below
the $^{12}$C/$^{13}$C and $^{16}$O/$^{18}$O abundance ratios by factors
of approximately 3 and 6, respectively. Thus, the non--detection of 
the $1_{10}-1_{11}$ H$_2^{13}$CO and H$_2$C$^{18}$O lines does not
contradict to the known values of the $^{12}$C/$^{13}$C and
$^{16}$O/$^{18}$O abundance ratios.

\subsection{Undetected lines}
Table~\ref{neg} presents the upper limits to the antenna temperatures of
observed but undetected lines together with the relevant column density limits.
The column density limits were calculated from Eq.(1) under the assumption
that the line populations are thermalized at 10~K. Most of the molecular
species from Table~\ref{neg} have already been found in TMC-1 at higher
frequencies.
Comparison of the published column densities with the upper limits
in Table~\ref{neg} shows that an increase in sensitivity by a factor
of 2--5 could lead to the detection of a much larger number of molecular
lines. Thus, molecular observations at low frequencies (4--10 GHz) may be
useful for studying dark clouds if the sensitivity (at $3\sigma$ level)
is about 5--10 mK or better.

We did not estimate H$_2^{13}$CO and H$_2$C$^{18}$O column densities
because of well-known strong deviations from LTE of the level populations
of these species.
Note that some other molecules may remain undetected in dark clouds due
to excitation particularities rather than low abundances. For instance, 
CH$_2$NH was not detected in TMC-1 and L134 by \cite{dic97}; they
suggested that it may be an excitation effect, since
the observed transitions have critical densities of about $10^6$~cm$^{-3}$,
which is well below the density of TMC-1.  
Our non--detection of the $1_{10}-1_{11}$ CH$_2$NH emission line is probably
a result of the $1_{10}-1_{01}$ transition at 167.4~GHz 
($E_{low}=2.135$~cm$^{-1}$), which can radiatively depopulate 
the $1_{10}$ level at the moderate densities found in TMC-1.

\section{Summary}
The results of the lowest frequency spectral survey
carried out toward a dark cloud, studying TMC-1 in the 4--6 GHz range using the
Arecibo radio telescope, and those from sensitive observations
at selected frequencies in  C-- and X--bands can be summarized
as follows:

\begin{itemize}
\item
A number of molecular lines were detected. The majority of the detected
lines belong to the cyanopolyynes HC$_3$N, HC$_5$N, HC$_7$N, and HC$_9$N.
\item
The rotational temperatures of the detected molecules fall in the range
4--9 K. Cyanopolyyne column densities vary from $5.6\times 10^{13}$~cm$^{-2}$ for
HC$_5$N to $2.7\times 10^{12}$~cm$^{-2}$ for HC$_9$N.
\item
Molecular observations at low frequencies (4--10 GHz) can be useful
for studying dark clouds, but to be really effective, the sensitivity must
be increased by a factor of 2 to 3. To date, observations at around 10~GHz
have been more fruitful than those at lower frequencies.

\end{itemize}

\acknowledgments

The National Astronomy and Ionosphere Center is operated by Cornell University under a
cooperative agreement with the National Science Foundation.
We are grateful to the staff of the Arecibo Observatory, and especially
Jeff Hagen, Mike Nolan, Phil Perillat, Chris Salter, and Arun
Venkataraman for help during the observations and useful discussions.
We thank the anonymous referee for important comments. SVK's visit
to Arecibo was supported by the NAIC Director's Office.
The work was partly supported by the Russian Foundation for Basic
Research (grant no. 01-02-16902).

\clearpage

\clearpage 

\clearpage

\clearpage

\clearpage

\clearpage

\clearpage

\begin{deluxetable}{llllllr}
\tabletypesize{\footnotesize}
\tablecaption{Gaussian fit parameters of the detected lines.
\label{gauss}} 
\tablewidth{0pt}
\tablehead{
\colhead{Molecule}&\colhead{Frequency}&\colhead{Transition}
        &\colhead{$\int T_A^*dV$}&\colhead{$V_{\rm LSR}$ }&
	  \colhead{$\Delta V$  }&\colhead{$T_A^*$}  \\
\colhead{}&\colhead{(MHz)} &\colhead{}&\colhead{K km s$^{-1}$ }
    &\colhead{km s$^{-1}$} &\colhead{km s$^{-1}$ }   & \colhead{(K)}    \\
}
\startdata     
HC$_3$N      &9097.0346 &$J=1-0\;F=1-1$      & 0.138(0.003)    &5.689(0.002) &0.167(0.001) &0.684\\
             &          &                    & 0.091(0.002)    &5.851(0.003) &0.168(0.006) &0.512\\
             &          &                    & 0.071(0.001)    &6.061(0.003) &0.167(0.007) &0.400\\
HC$_3$N      &9089.3321 &$J=1-0\;F=2-1$      & 0.180(0.012)    &5.689(0.002) &0.167(0.001) &1.085\\
             &          &                    & 0.136(0.007)    &5.851(0.003) &0.168(0.006) &0.814\\
             &          &                    & 0.115(0.003)    &6.061(0.003) &0.167(0.007) &0.693\\
HC$_3$N      &9100.2727 &$J=1-0\;F=0-1$      & 0.042(0.004)    &5.689(0.002) &0.167(0.001) &0.240\\
             &          &                    & 0.032(0.003)    &5.851(0.003) &0.168(0.006) &0.180\\
             &          &                    & 0.024(0.001)    &6.061(0.003) &0.167(0.007) &0.137\\
H$^{13}$CCCN &8817.096  &$J=1-0$ $F=2-1$     &0.006(0.002)     &5.66(0.14)  &0.40(0.00)\tablenotemark{1} &0.014  \\
HC$^{13}$CCN &9059.736  &$J=1-0$ $F=2-1$     &0.007(0.002)     &5.91(0.06)  &0.40(0.00)\tablenotemark{1} &0.015  \\
HCC$^{13}$CN &9060.608  &$J=1-0$ $F=2-1$     &0.009(0.002)     &5.72(0.04)  &0.40(0.00)\tablenotemark{1} &0.021  \\
HCC$^{13}$CN &9059.318  &$J=1-0$ $F=1-1$     &0.008(0.002)     &6.11(0.15)  &0.40(0.00)\tablenotemark{1} &0.018  \\
HC$_5$N      &5325.421  &$J=2-1$ $F=3-2$     &0.12(0.003)      &5.91(0.01)  &0.41(0.01)     &0.28  \\
HC$_5$N      &5325.330  &$J=2-1$ $F=2-1$     &0.06(0.002)      &5.91(0.01)  &0.41(0.01)     &0.15  \\
HC$_5$N      &5324.270  &$J=2-1$ $F=1-0$     &0.03(0.001)      &5.91(0.01)  &0.41(0.01)     &0.07  \\
HC$_7$N*    &5640.0316  &$J=5-4\; F=6-5$     &0.027(0.004)     &6.00(0.02)  &0.34(0.04)     &0.075 \\
HC$_7$N*    &5640.0091  &$J=5-4\; F=5-4$     &0.021(0.004)     &6.00(0.02)  &0.34(0.04)     &0.061 \\
HC$_7$N*    &5639.9580  &$J=5-4\; F=4-3$     &0.018(0.004)     &6.00(0.02)  &0.34(0.04)     &0.049 \\
HC$_7$N     &7896.023   &$J=7-6\; F=8-7$     &0.028(0.012)     &5.92(0.04)  &0.33(0.04)     &0.080 \\
HC$_7$N     &7896.010   &$J=7-6\; F=7-6$     &0.024(0.012)     &5.92(0.04)  &0.33(0.04)     &0.069 \\
HC$_7$N     &7895.989   &$J=7-6\; F=6-5$     &0.021(0.012)     &5.92(0.04)  &0.33(0.04)     &0.060 \\
HC$_7$N     &9024.020   &$J=8-7\; F=9-8$     &0.028(0.006)     &5.94(0.02)  &0.39(0.03)     &0.069 \\
HC$_7$N     &9024.009   &$J=8-7\; F=8-7$     &0.025(0.006)     &5.94(0.02)  &0.39(0.03)     &0.061 \\
HC$_7$N     &9023.994   &$J=8-7\; F=7-6$     &0.022(0.006)     &5.94(0.02)  &0.39(0.03)     &0.053 \\
HC$_9$N*    &5229.3272  &$J=9-8$\tablenotemark{2}
                                             &0.016(0.004)\tablenotemark{3}
                                                               &5.87(0.14)  &1.11(0.23) &0.014 \\
HC$_9$N*    &5810.362   &$J=10-9$\tablenotemark{2}
                                             &0.017(0.005)\tablenotemark{3}
                                                               &6.30(0.23)  &1.51(0.39) &0.012 \\
HC$_9$N*    &8134.503   &$J=14-13$\tablenotemark{2}
                                             &0.016(0.005)\tablenotemark{3}
					                       &5.90(0.11)  &0.57(0.21)     &0.027 \\
HC$_9$N*    &8715.538   &$J=15-14$\tablenotemark{2}
                                             &0.022(0.003)     &5.96(0.02)  &0.39(0.05)     &0.053 \\
HC$_9$N*    &9296.5721  &$J=16-15$\tablenotemark{2}
                                             &0.022(0.003)     &5.91(0.03)  &0.33(0.06)     &0.061 \\
CCS*       &5402.6175 &$N_J=3_2-2_2$         &0.009(0.003)\tablenotemark{3}
                                                               &5.70(0.07)  &0.40(0.00)\tablenotemark{1}
                                                                                           &0.021 \\
C$_3$S*    &5780.759   &$J=1-0$             &0.041(0.005)      &5.91(0.01)  &0.49(0.03)     &0.077 \\

C$_4$H     &9497.616   &$N=1-0 \; J=3/2-1/2$ &0.098(0.01)      &5.77(0.01)  &0.21(0.03)     &0.447 \\
           &            &            $F=2-1$ &                 &            &               &      \\
C$_4$H$_2$* &8931.92    &$1_{01}-0_{00}$     &0.009(0.003)\tablenotemark{3}
                                                               &5.72(0.01)  &0.11(0.30)     &0.079 \\
H$_2$CO    &4829.6594 &$1_{10}-1_{11}$ $F=2-2$\tablenotemark{4}
                                            &$-0.533$          &5.84(0.01)  &0.71(0.02)     &$-0.753$\\
%
\enddata

\tablenotetext{1}{the line width is fixed}
\tablenotetext{2}{the sum of the hyperfine components presented in Fig.~4}
\tablenotetext{3}{marginal detection}
\tablenotetext{4}{the strongest hyperfine component}

\tablecomments{The frequencies
of the known lines are taken from the Lovas catalog of observed lines.
The frequences of the newly detected lines (denoted by an asterisk after
the species name) and all HC$_7$N lines are taken from the Cologne database
for molecular spectroscopy. The numbers in parentheses denote the $1\sigma$
uncertainties of the corresponding values.}
\end{deluxetable}

\clearpage

\begin{deluxetable}{lcc}
\tablecaption{Rotational temperatures and column densities 
for detected molecular species.
\label{trot}}
\tablewidth{0pt}
\tabcolsep=30pt
\tablehead{
\colhead{Molecule} &\colhead{$T_{\rm rot}$}&\colhead{Column density}\\
\colhead{}   &\colhead{  (K)}   & \colhead{(cm$^{-2}$)} \\
}
\startdata
HC$_5$N      &  $4.3\pm 0.4$        &$(5.6\pm 1.4)\times 10^{13}$\\
HC$_7$N      &  $7.7\pm 0.7$        &$(1.2\pm 0.2)\times 10^{13}$\\  
HC$_9$N      &  $8.4\pm 0.4$        &$(2.7\pm 0.2)\times 10^{12}$\\
CCS          &  $5.7\pm 1.0$        &$(3.4\pm 0.8)\times 10^{13}$\\
C$_4$H       & 5.6\tablenotemark{1} &$1.1\times 10^{15}$\\
C$_4$H$_2$   & 4.2\tablenotemark{2} &$4.2\times 10^{12}$\\
\enddata

\tablenotetext{1}{from \cite{ohi98}.}

\tablenotetext{2}{from \cite{kaw91}.}

\tablecomments{Errors are given at $1\sigma$ level. For cyanopolyynes and CCS
the parameters were obtained by using rotation diagrams. For C$_4$H and
C$_4$H$_2$, the column densities were calculated using rotation temperatures
taken from the literature.}

\end{deluxetable}

\clearpage

\begin{deluxetable}{lllccc}
\tablecaption{Upper limits to the antenna temperatures for undetected lines
and relevant molecular column density limits.
\label{neg}}
\tablewidth{0pt}
\tablehead{
\colhead{Molecule} &\colhead{Frequency}&\colhead{Transition} &
        \colhead{$|T_A^*|$ upper} &\colhead{Column density\tablenotemark{1} } &Reference for\\
\colhead{}&\colhead{(MHz)}&\colhead{}&\colhead{limit (K)}&\colhead{upper limit (cm$^{-2}$)}
                                                                &frequency\tablenotemark{2}
}
\startdata     			           
HC$_9$N\tablenotemark{3}
                &4067.2695\tablenotemark{4}
                           &$J=7-6$                  &0.026 &                       & Col\\
HC$_{11}$N      &5071.8851 & $J=15-14$               &0.027 &  $2.53\times 10^{12}$ & JPL\\
HC$_{11}$N      &5410.0103 & $J=16-15$               &0.012 &  $1.01\times 10^{12}$ & JPL\\
HC$_{11}$N      &8115.007  & $J=24-23$               &0.033 &  $2.41\times 10^{12}$ & JPL\\
HC$_{11}$N      &8791.257  & $J=26-25$               &0.042 &  $3.12\times 10^{12}$ & JPL\\
HC$_{11}$N      &9129.376  & $J=27-26$               &0.018 &  $1.29\times 10^{12}$ & JPL\\
HC$_{13}$N      &5348.622  & $J=25-24$               &0.012 &  $1.33\times 10^{12}$ & Col\\
HC$_{13}$N      &5562.566  & $J=26-25$               &0.013 &  $1.26\times 10^{12}$ & Col\\
C$_4$D          &8868.9005 &$N=1-0$                  &0.033 &  $2.25\times 10^{14}$ & Col\\ 
                &          &$J=1/2-1/2$              &      &                       &    \\ 
C$_5$O          &5467.388  & $J=2-1$                 &0.030 &  $5.00\times 10^{12}$ & JPL\\ 
C$_4$S          &5912.1755 &$N_J=3_2-2_1$            &0.027 &  $1.11\times 10^{13}$ & Col\\ 
C$_4$S          &8868.7075 & $N_J=4_3-3_2$           &0.033 &  $1.04\times 10^{13}$ & Col\\ 
C$_5$S          &5536.222  & $J=3-2$                 &0.029 &  $4.43\times 10^{12}$ & JPL\\ 
C$_5$S          &9227.028  & $J=5-4$                 &0.036 &  $3.81\times 10^{12}$ & JPL\\ 
C$_5$N  &5607.1377\tablenotemark{4} &$N=2-1$         &0.030 &  $2.56\times 10^{13}$ & Col\\
                &                   &$J=5/2-3/2$     &      &                       &    \\		     
C$_6$H &4201\tablenotemark{5}&$^2\Pi_{1/2}$ $J=3/2-1/2$&0.026& $1.86\times 10^{13}$ & JPL\\ 
CH$_3$C$_4$H    &4071.4935 & $J_K=1_0-0_0$           &0.030 &  $1.36\times 10^{14}$ & Col\\ 
CH$_2$NH &5289.813\tablenotemark{4}&$1_{10}-1_{11}$  &0.030 &  $3.54\times 10^{13}$ & Lov\\  
HCOOH           &4916.312  &$2_{11}-2_{12}$          &0.039 &  $9.63\times 10^{13}$ & Lov\\ 
CH$_2$CO        &5660.9476 & $5_{14}-5_{15}$         &0.024 &  $3.68\times 10^{14}$ & JPL\\ 
H$_2$C$^{18}$O  &4388.8011\tablenotemark{4} &$1_{10}-1_{11}$ &0.021 &  &Lov\\ 
H$_2^{13}$CO    &4593.0865\tablenotemark{4} &$1_{10}-1_{11}$ &0.072 &  &Lov\\ 

\enddata

\tablenotetext{1}{assuming a rotational temperature equal to 10 K}

\tablenotetext{2}{JPL indicates the JPL catalog, Col the Cologne database, 
and Lov the Lovas line list}

\tablenotetext{3}{HC$_9$N column density is presented in Table 2.}

\tablenotetext{4}{blended hyperfine structure}

\tablenotetext{5}{a line multiplet}

\tablecomments{The limits are given at $3\sigma$ level.}
\end{deluxetable}

\end{document}